\documentclass[
]{ceurart}

\usepackage{listings}
\begin{document}

\conference{In: 
B. Combemale, G. Mussbacher, S. Betz, A. Friday, I. Hadar, J. Sallou, I. Groher, H. Muccini,
O. Le Meur, C. Herglotz, E. Eriksson, B. Penzenstadler, AK. Peters, C. C. Venters. 
Joint Proceedings of ICT4S 2023 Doctoral Symposium, Demonstrations \& Posters Track and Workshops. Co-located with ICT4S 2023. Rennes, France, June 05-09, 2023.}

\copyrightyear{2023}
\copyrightclause{Copyright for this paper by its authors.
  Use permitted under Creative Commons License Attribution 4.0
  International (CC BY 4.0).}

\title{Scale-Score: Food Label to Support Nutritious and Sustainable Online Grocery Shopping - Extended Abstract}

\author[1]{Marco Druschba}[%
orcid=0000-0002-7634-9482,
email=marco-druschba@web.de,
]

\author[2]{Gözel Shakeri}[%
orcid=0000-0002-3154-0814,
email=gozel.shakeri@uol.de,
]

\address{Carl von Ossietzky University of Oldenburg, Germany}

\cormark[2]

\begin{abstract}
  To empower online grocery shoppers in making nutritionally and environmentally informed decisions, we investigate the efficacy of the Scale-Score, a label combining nutritional and environmental information to highlight a product's benefit to both the consumer's and the planet's health, without obscuring either information. We conducted an experimental study in a mock online grocery environment, and assessed label efficacy. We find that the Scale-Score supports nutritious purchases, yet needs improving regarding sustainability support. Our research shows first insights into design considerations and performance of a combined yet disjoint food label. 
\end{abstract}

\begin{keywords}
  Sustainable HCI \sep
  persuasive technology \sep
  sustainability communication \sep
  personal informatics
\end{keywords}

\maketitle

\section{Introduction}

Labels support consumers in making nutritious and sustainable decisions by transforming complex information about food e.g., nutritional values, animal welfare standards, or environmental aspects, into simple logos or diagrams \cite{lemken2021climatescorelabel}. Food-label technology and personal informatics thereby both use similar techniques to motivate users \cite{sauve2020econundrum}, such as providing information, enabling comparison, and giving feedback \cite{Froehlich2010ecotechnology}. 
Several studies within the HCI discipline investigated labels as a means of providing education tailored to users' own context and choices; addressing health- and environmental challenges separately, although they are closely intertwined (e.g. sustainability: Envirofy \cite{shakeri2021envirofy}, Nu-Food \cite{PANZONE2021nufood}; nutrition: BetterChoice \cite{fuchs2022betterchoice}, FLICC \cite{Harrington2019Flicc}). 

\begin{figure}
    \centering
    \includegraphics[width=\textwidth]{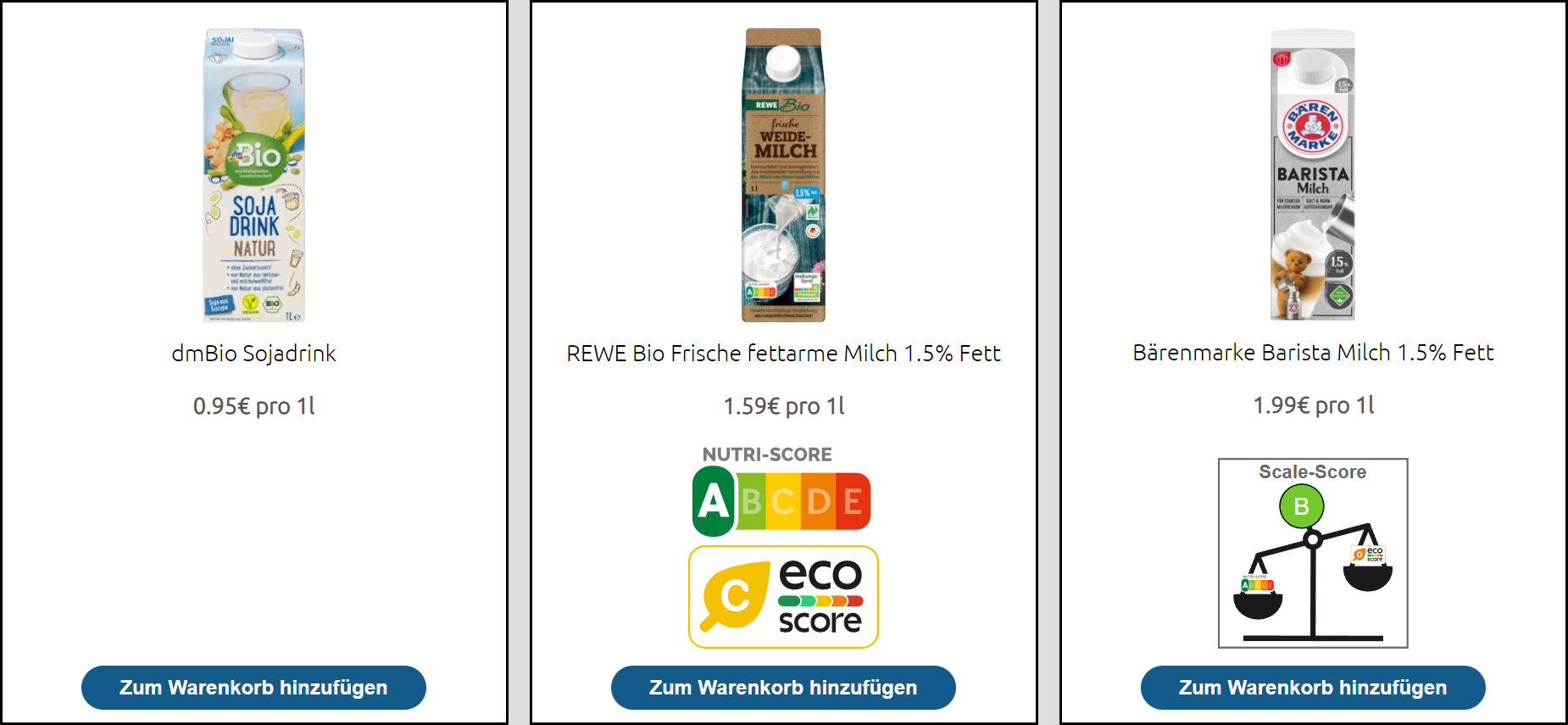}
    \caption{Scale-Score (right) provides high level information about the nutritional and sustainable value of foods, yet gives additional information to allow for individual prioritisation of health or environment. In a mock-up online supermarket we tested three conditions (left: no labels, middle: Nutri- and Eco-Score, right: Scale-Score) showing that the Scale-Score supports nutritious purchase decisions, yet needs improving regarding sustainability support.}
    \label{fig:all}
\end{figure}

Our research focuses on the design space of labels which comprise of both, health and environmental information; when it matters most, \textit{while} online grocery shopping \cite{zapico2016ecopanel, Luo2017_OnlineEngagementWebbaseSupermarketHealthProgram}. We describe a study which tested the impact of presenting the \textit{Scale-Score} (Figure \ref{fig:all}, right) on the nutritional quality and environmental impact of the consumers' food choices, compared to the effects of both Nutri-Score and Eco-Score labels, and no persuasive technology. 
We found the Scale-Score improved nutritional quality of purchases, however surprisingly, it performed worse in terms of environmental impact, compared to Nutri-Eco and baseline condition. 
This paper contributes first evidence in support of using a joint yet disjoint nutritional and ecological label to encourage transitions towards healthier and more sustainable diets, when online shopping. 

\section{Experimental Research}

\subsection{Scale-Score}

The Scale-Score (Figure \ref{fig:all}, right) combines Nutri- and Eco-Scores into a single label represented by a classic beam scale. It also shows an overall rating based on the product's Nutri- and Eco-Scores by computing their mean value; in case of an uneven result, we opted to go in favour of the Nutri-Score, prioritising nutrition over sustainability, in accordance with previously gathered user requirements. 

\subsection{Methods \& Participants}
We employed a within-subjects design with a single independent variable, visualisation, and three factors, Scale-Score, Nutri-/Eco-Score, and baseline with no visualisation. Dependent variables (i.e., shopping behaviour) were: 1) average environmental value of chosen products (based on Eco-Score calculations) and 2) average nutrition value of chosen products (based on Nutri-Score calculations). In each condition, participants shopped according to a shopping list with three items (cereal, milk, and peanut butter). Participants were entered into a random draw to win their shopping basket, as an incentive to encourage normal purchase decisions. The study lasted one hour.

We recruited 12 participants (5f, $\mu =$ 39 years, $\sigma =$ 22.9 years) through our institution's forums. Regarding demographics, all participants stated having seen the Nutri-Score prior; the Eco-Score was seen by 16.7\,\% prior. 

\subsection{Results}

We used a one-way ANOVA with post-hoc Tukey-tests via IBM SPSS Statistics (v. 28.0.1.0). For this study an alpha ($\alpha$) of 0.05 was used. 

There was no statistically significant difference between the conditions on `nutrition' as determined by one-way ANOVA $(F(2.35) = 0.8 p = 0.458)$. However, a Tukey-post-hoc-test revealed that Scale- and Nutri-/Eco-Score means did not differ from each other ($p$-value $=0.994$), but both Nutri-/Eco-Score ($p$-value $=0.557$) and Scale-Score ($p$-value $=0.496$) differed compared to no score. Looking at the descriptive statistics, Scale-Score resulted in lower nutrition values (mean $=2.89$) compared to Nutri-/Eco-Score (mean $= 3.06$) and no score (mean $= 4.78$).

There also was no statistically significant difference between the conditions on `sustainability' as determined by one-way ANOVA $(F(2.35) = 1,301 p = 0.286)$. Tukey-post-hoc-test revealed a non-significant difference between Nutri-/Eco-Score ($p$-value $=0.595$) and Scale-Score ($p$-value $=0.810$) compared to no score. Looking at the descriptive statistics, Scale-Score resulted in lowest sustainability values (mean $=53.11$), Nutri-/Eco-Score in highest (mean $=59.78$) and no score in intermediate (mean $=55.69$).

\section{Discussion and Conclusions}

To achieve a global and successful transition to healthy and sustainable diets, systems and tools are needed to support consumers in this. We designed Scale-Score, a label that displays nutritional and environmental information. 

We did not find significant differences in support provision of either visualisation compared to baseline, however there is a trend showing the Nutri- and Eco-Score combination may support consumers in sustainable and healthy decision making. The Scale-Score may support nutritious choices compared to no visualisation, however, worsened environmental impact of the basket compared to baseline. First, this may be due to the make-up of the Scale-Score: nutritional aspect weighted more into the final score. Consequently, a product that is marked with a good Scale-Score rating (e.g. B) may well contain an environmental `D' rating. As a result, the average sustainability score was worse compared to Nutri- and Eco-Score representation, resulting in Scale-Score's poor environmental performance. Second, participants may have ignored the multi-level information provided, given the small sizes of Nutri- and Eco-Score labels within the Scale-Score, contributing further to the de-valuation of environmental information. 

In future, we plan to re-design the label taking advantage of the interaction modalities available in web-based interfaces, where meta \textit{and} multi-level information can effectively support sustainable and nutritious grocery shopping. At ICT4S, we hope to inspire and engage in conversations on user-centred food label designs embedded in personal informatics.  

\newpage
\bibliography{main}

\begin{thebibliography}{9}
\expandafter\ifx\csname natexlab\endcsname\relax\def\natexlab#1{#1}\fi
\providecommand{\url}[1]{\texttt{#1}}
\providecommand{\href}[2]{#2}
\providecommand{\path}[1]{#1}
\providecommand{\DOIprefix}{doi:}
\providecommand{\ArXivprefix}{arXiv:}
\providecommand{\URLprefix}{URL: }
\providecommand{\Pubmedprefix}{pmid:}
\providecommand{\doi}[1]{\href{http://dx.doi.org/#1}{\path{#1}}}
\providecommand{\Pubmed}[1]{\href{pmid:#1}{\path{#1}}}
\providecommand{\bibinfo}[2]{#2}
\ifx\xfnm\relax \def\xfnm[#1]{\unskip,\space#1}\fi
\bibitem[{Lemken et~al.(2021)Lemken, Zühlsdorf, and
  Spiller}]{lemken2021climatescorelabel}
\bibinfo{author}{D.~Lemken}, \bibinfo{author}{A.~Zühlsdorf},
  \bibinfo{author}{A.~Spiller},
\newblock \bibinfo{title}{Improving consumers’ understanding and use of
  carbon footprint labels on food: Proposal for a climate score label},
\newblock \bibinfo{journal}{EuroChoices} \bibinfo{volume}{20}
  (\bibinfo{year}{2021}) \bibinfo{pages}{23--29}.
  \DOIprefix\doi{https://doi.org/10.1111/1746-692X.12321}.
\bibitem[{Sauv\'{e} et~al.(2020)Sauv\'{e}, Bakker, and
  Houben}]{sauve2020econundrum}
\bibinfo{author}{K.~Sauv\'{e}}, \bibinfo{author}{S.~Bakker},
  \bibinfo{author}{S.~Houben},
\newblock \bibinfo{title}{Econundrum: Visualizing the climate impact of dietary
  choice through a shared data sculpture},
\newblock in: \bibinfo{booktitle}{Proceedings of the 2020 ACM Designing
  Interactive Systems Conference}, DIS '20, \bibinfo{publisher}{Association for
  Computing Machinery}, \bibinfo{address}{New York, NY, USA},
  \bibinfo{year}{2020}, p. \bibinfo{pages}{1287–1300}.
  \DOIprefix\doi{10.1145/3357236.3395509}.
\bibitem[{Froehlich et~al.(2010)Froehlich, Findlater, and
  Landay}]{Froehlich2010ecotechnology}
\bibinfo{author}{J.~Froehlich}, \bibinfo{author}{L.~Findlater},
  \bibinfo{author}{J.~Landay},
\newblock \bibinfo{title}{The design of eco-feedback technology},
\newblock in: \bibinfo{booktitle}{Proceedings of the SIGCHI Conference on Human
  Factors in Computing Systems}, CHI '10, \bibinfo{publisher}{Association for
  Computing Machinery}, \bibinfo{address}{New York, NY, USA},
  \bibinfo{year}{2010}, p. \bibinfo{pages}{1999–2008}.
  \DOIprefix\doi{10.1145/1753326.1753629}.
\bibitem[{Shakeri and McCallum(2021)}]{shakeri2021envirofy}
\bibinfo{author}{G.~Shakeri}, \bibinfo{author}{C.~H. McCallum},
\newblock \bibinfo{title}{Envirofy your shop: Development of a real-time tool
  to support eco-friendly food purchases online},
\newblock in: \bibinfo{booktitle}{Extended Abstracts of the 2021 CHI Conference
  on Human Factors in Computing Systems}, CHI EA '21,
  \bibinfo{publisher}{Association for Computing Machinery},
  \bibinfo{address}{New York, NY, USA}, \bibinfo{year}{2021}.
  \DOIprefix\doi{10.1145/3411763.3451713}.
\bibitem[{Panzone et~al.(2021)Panzone, Ulph, Zizzo, Hilton, and
  Clear}]{PANZONE2021nufood}
\bibinfo{author}{L.~A. Panzone}, \bibinfo{author}{A.~Ulph},
  \bibinfo{author}{D.~J. Zizzo}, \bibinfo{author}{D.~Hilton},
  \bibinfo{author}{A.~Clear},
\newblock \bibinfo{title}{The impact of environmental recall and carbon
  taxation on the carbon footprint of supermarket shopping},
\newblock \bibinfo{journal}{Journal of Environmental Economics and Management}
  \bibinfo{volume}{109} (\bibinfo{year}{2021}) \bibinfo{pages}{102137}.
  \DOIprefix\doi{10.1016/j.jeem.2018.06.002}.
\bibitem[{Fuchs et~al.(2022)Fuchs, Lian, Michels, Mayer, Toniato, and
  Tiefenbeck}]{fuchs2022betterchoice}
\bibinfo{author}{K.~L. Fuchs}, \bibinfo{author}{J.~Lian},
  \bibinfo{author}{L.~Michels}, \bibinfo{author}{S.~Mayer},
  \bibinfo{author}{E.~Toniato}, \bibinfo{author}{V.~Tiefenbeck},
\newblock \bibinfo{title}{Effects of digital food labels on healthy food
  choices in online grocery shopping},
\newblock \bibinfo{journal}{Nutrients} \bibinfo{volume}{14}
  (\bibinfo{year}{2022}). \DOIprefix\doi{10.3390/nu14102044}.
\bibitem[{Harrington et~al.(2019)Harrington, Scarborough, Hodgkins, Raats,
  Cowburn, Dean, Doherty, Foster, Juszczak, Ni~Mhurchu, Winstone, Shepherd,
  Timotijevic, and Rayner}]{Harrington2019Flicc}
\bibinfo{author}{R.~A. Harrington}, \bibinfo{author}{P.~Scarborough},
  \bibinfo{author}{C.~Hodgkins}, \bibinfo{author}{M.~M. Raats},
  \bibinfo{author}{G.~Cowburn}, \bibinfo{author}{M.~Dean},
  \bibinfo{author}{A.~Doherty}, \bibinfo{author}{C.~Foster},
  \bibinfo{author}{E.~Juszczak}, \bibinfo{author}{C.~Ni~Mhurchu},
  \bibinfo{author}{N.~Winstone}, \bibinfo{author}{R.~Shepherd},
  \bibinfo{author}{L.~Timotijevic}, \bibinfo{author}{M.~Rayner},
\newblock \bibinfo{title}{A pilot randomized controlled trial of a digital
  intervention aimed at improving food purchasing behavior: The front-of-pack
  food labels impact on consumer choice study},
\newblock \bibinfo{journal}{JMIR Form Res} \bibinfo{volume}{3}
  (\bibinfo{year}{2019}) \bibinfo{pages}{e9910}.
  \DOIprefix\doi{10.2196/formative.9910}.
\bibitem[{Zapico et~al.(2016)Zapico, Katzeff, Bohn\'{e}, and
  Milestad}]{zapico2016ecopanel}
\bibinfo{author}{J.~L. Zapico}, \bibinfo{author}{C.~Katzeff},
  \bibinfo{author}{U.~Bohn\'{e}}, \bibinfo{author}{R.~Milestad},
\newblock \bibinfo{title}{Eco-feedback visualization for closing the gap of
  organic food consumption},
\newblock in: \bibinfo{booktitle}{Proceedings of the 9th Nordic Conference on
  Human-Computer Interaction}, NordiCHI '16, \bibinfo{publisher}{Association
  for Computing Machinery}, \bibinfo{address}{New York, NY, USA},
  \bibinfo{year}{2016}. \DOIprefix\doi{10.1145/2971485.2971507}.
\bibitem[{Luo et~al.(2017)Luo, Li, Berkovsky, Koprinska, and
  Chen}]{Luo2017_OnlineEngagementWebbaseSupermarketHealthProgram}
\bibinfo{author}{L.~Luo}, \bibinfo{author}{B.~Li},
  \bibinfo{author}{S.~Berkovsky}, \bibinfo{author}{I.~Koprinska},
  \bibinfo{author}{F.~Chen},
\newblock \bibinfo{title}{Online engagement for a healthier you: A case study
  of web-based supermarket health program},
\newblock in: \bibinfo{booktitle}{Proceedings of the 26th International
  Conference on World Wide Web Companion}, WWW '17 Companion,
  \bibinfo{publisher}{International World Wide Web Conferences Steering
  Committee}, \bibinfo{address}{Republic and Canton of Geneva, CHE},
  \bibinfo{year}{2017}, p. \bibinfo{pages}{1053–1061}.
  \DOIprefix\doi{10.1145/3041021.3055129}.

\end{thebibliography}

\appendix

\section{Online Resources}

Longer version of this submission is available on \href{http://oops.uni-oldenburg.de/5512/}{http://oops.uni-oldenburg.de/5512} (German). Source code for the online supermarket is available via \href{https://github.com/md812/groceries}{GitHub}.

\end{document}